\documentclass[prb,aps,twocolumn,showpacs,bibnotes,superscriptaddress,epsf]{revtex4}

\usepackage{graphicx}
\usepackage{amsmath}
\usepackage{amssymb}
\usepackage{color}
\usepackage{dcolumn}
\usepackage{epsfig}
\usepackage{bm}
\usepackage{subfigure}
\usepackage[urlcolor=blue]{hyperref}
\hypersetup{backref, colorlinks=true, linkcolor=blue, citecolor=blue}

\begin{document}

\title{Observation of Meissner effect in potassium-doped $p$-quaterphenyl}

\author{Jia-Feng Yan}
\affiliation{Center for High Pressure Science and Technology Advanced Research, Shanghai 201203, China}

\author{Ren-Shu Wang}
\affiliation{Center for High Pressure Science and Technology Advanced Research, Shanghai 201203, China}
\affiliation{School of Materials Science and Engineering, Hubei University, Wuhan 430062, China}

\author{Kai Zhang}
\affiliation{Center for High Pressure Science and Technology Advanced Research, Shanghai 201203, China}

\author{Xiao-Jia Chen}
\email{xjchen@hpstar.ac.cn}
\affiliation{Center for High Pressure Science and Technology Advanced Research, Shanghai 201203, China}

\date{\today}

\begin{abstract}
Inspired by the discovery of high temperature superconductivity in potassium-doped $p$-terphenyl, we examine the possibility of superconductivity in $p$-quaterphenyl with one more phenyl ring than $p$-terphenyl. The Meissner effect with critical temperatures ranging from 3.5 K to 120 K is found by the magnetic susceptibility measurements in $p$-quaterphenyl upon doping potassium in the condition of annealing or just pestling. The primary superconducting phase with critical temperature of 7.2 K can be reduplicated in several superconducting samples. In both the annealed and pestled superconducting samples, the observation of bipolaronic character by the Raman scattering measurements reveals the close relationship between bipolarons and superconductivity. The occurrence of superconductivity in potassium-doped $p$-quaterphenyl provides an indication that chain link organic molecules are potential candidates for high temperature superconductors.
\end{abstract}
\pacs{74.70.-b, 74.20.Mn, 82.35.Lr, 78.30.Jw}

\maketitle

\section{INTRODUCTION}

The perpetual enthusiasm of chasing novel high temperature superconductors is never dampened. Organic compounds are predicted as candidates of superconductors with high critical temperature ($T$$_c$) even above room temperature.\cite{predict1,predict2} Poly($para$-phenylenes)(PPPs), as an important component of quasi-one-dimensional conducting polymers\cite{polymer}, have a great potential to be superconductors due to the great features of high conductivity,\cite{PPP1} the absence of isomerization upon doping, nondegenerate ground state structure, large Coulomb correlation,\cite{PPP2} and bipolaron conducting mechanism.\cite{PPP3} $p$-Oligophenyls with a shorter chain possess the similar molecular structure of PPP which with infinite benzene rings linked by single carbon-carbon bond in $para$ position. In fact, PPP also exhibits high electrical conductivity upon doping donors or acceptors.\cite{conduct} The conductivity of doped PPP significantly increases with the increasing chain length\cite{upconduct} in contrast to that the band gaps decrease as a function of increasing quinoid character of the backbone.\cite{quinoid}

Recently, potassium-doped $p$-terphenyl, as a member of $p$-oligophenyls, attracts a lot of attentions due to the discovery of superconductivity with critical temperatures of 7.2 K, 43 K and even 123 K.\cite{terphenyl1, terphenyl2, terphenyl3} These discoveries support previous predictions\cite{benzene} upon calculating which claim that potassium-doped solid benzene shows spuerconductivity with most stable phase of $T$$_c$ around 6.2 K and this superconducting phase shares the common feature of aromatic hydrocarbons superconductors. By comparing the first-principles calculations and the experimentally observed X-ray diffraction patterns,\cite{structure} the superconducting phase with $T$$_c$ of 7.2 K\cite{terphenyl1} in potassium-doped $p$-terphenyl should be a mixture phase with doping level of range from 2 to 3. The 7.2 K superconducting phase reemerges in potassium-doped $p$-quinquephenyl\cite{quinquephenyl} and 2,2$'$-bipyridine,\cite{bipyridine} respectively. Subsequently, a new report\cite{ARPES} provided the spectroscopic evidence for superconducting pairing gaps persisting to 60 K or above on the potassium surface-doped $p$-terphenyl single crystals by measurements of angle-resolved photoemission spectroscopy (ARPES). The gapped phases were also observed in K-doped single layer $p$-terphenyl films grown on Au (111).\cite{FengDL} The unresponsive feature of this gap to the applied magnetic field up to 11 Tesla implies the extremely high upper critical field for such high $T$$_c$ superconductors. Afterwards, another superconducting phase with high $T$$_c$ of 107 K was reported in $p$-terphenyl flakes upon doping potassium.\cite{india} The discussion about band structure and density of states of $p$-terphenyl helps to search high $T$$_c$ superconductors with similar electronic structure to $p$-terphenyl.\cite{DOS} A proposal of the mechanism driving the remarkable superconducting transition in K-doped $p$-terphenyl is the Fano resonance between superconducting gaps near a Lifshitz transition.\cite{fano} Meanwhile, a superconductivity-like transition was observed in potassium-doped $p$-terphenyl and $p$-quaterphenyl by high-pressure synthesis.\cite{WEN} The observations of ferromagnetic background and the absence of diamagnetic magnetic susceptibility make them to claim that it is insufficient to regard potassium-doped $p$-terphenyl and $p$-quaterphenyl as superconductors. Superconductivity in potassium-doped $p$-terphenyl gets supports from both experimental\cite{terphenyl1,terphenyl2,terphenyl3,ARPES,india,FengDL} and theoretical\cite{structure,DOS,fano} works. Thus, exploring the possibility of superconductivity in $p$-quaterphenyl which consists one more benzene ring than $p$-terphenyl is worthy of further exploration.

$p$-Quaterphrnyl, as a member of conjugated semiconducting oligomers, has an extensive potential for organic thin film transistors,\cite{OTFT} organic light emitting diodes (OLED),\cite{OLED} and organic lasers\cite{laser} due to the optical activation.\cite{optical} Oriented $p$-quaterphenyl layers show favorable feature of charge carrier transport in comparison with bulk material.\cite{carrier} Such $\pi$-conjugated oligomers like $p$-quaterphenyl thin films express an obvious trend towards crystallize,\cite{crystal} and the thin films of crystalline organic semiconductors exhibit anisotropy in charge transport. Simultaneously, depending on the conjugated structure, the derivatives of $p$-quaterphenyl also show strong blue light emission in solution.\cite{derivative} Both these optical and electronic features make the $p$-quaterphenyl thin films and derivatives available for photoelectric devices. The absolute advantages of organic over conventional thin film transistor materials, such as the permit of mechanically rugged and flexible systems,\cite{OTFT} drive people to create organic thin film materials with high quality under the influence of temperature,\cite{anneal,temperature} magnetic field,\cite{magnet} and surfactant.\cite{pattern} After the applications in photoelectricity, the high possibility of superconductivity in $p$-quaterphenyl will enrich other applications.

In this work, we report the finding of superconductivity in potassium-doped $p$-quaterphenyl by annealing or pestling process. The magnetization measurements provide the solid evidences of the presence of Meissner effect with $T$$_c$$'$s ranging from 3.5 K to 120 K. The primary 7.2 K superconducting phase can be reduplicated in both the annealed and pestled samples. Another weak superconducting phase with a high $T$$_c$ of 120 K is observed in the pestled sample as well. The realization of superconductivity in potassium-doped $p$-quaterphenyl adds a new leaguer of superconductors in $p$-oligophenyl family. The Raman scattering measurements provide the evidences of the formation of bipolarons. Otherwise, this work also provides a simple method to synthesize the organic superconductors by pestling without annealing.

\section{EXPERIMENTAL DETAILS}

We synthesized samples by doping potassium into $p$-quaterphenyl. The white powdery $p$-quaterphenyl (99.5$\%$, purity) and potassium metal (99$\%$, purity) were purchased from Alfa Aesar and Sinopharm Chemical Reagent, respectively. In an argon atmosphere of glove box with the oxygen and moisture levels less than 0.1 ppm, potassium was cut into pieces and mixed with $p$-quaterphenyl with the mole ratio 3:1$\sim$4:1. Then the mixtures were sealed in quartz tubes under high vacuum (1$\times$10$^{-4}$Pa) and heated at temperature 605$\sim$625 K for 72$\sim$240 hours in furnace. After the annealing, the formerly white powder turned to black solid particles. The annealed sample tubes were opened in glove box and the samples were pestled into powder in an agate mortar. In addition, we also prepared samples by pestling the mixtures of $p$-quaterphenyl and potassium without annealing. The powdery mixture gradually turned to black by pestling in an agate mortar. Both annealed and pestled samples were loaded into non-magnetic capsules and capillary tubes for following magnetization and Raman scattering measurements. The magnetization measurements were performed with a SQUID magnetometer (Quantum Design MPMS3) in the temperature range of 1.8$\sim$300 K. An in-house system with Charge Coupled Device and Spectrometer  Princeton Instruments in a wavelength of 660 nm was used to collect Raman scattering spectra.

\section{RESULTS AND DISCUSSION}

\begin{table}[tbp]
\centering
\caption{Summary of K-doped \emph{p}-quaterphenyl samples with different superconducting transition temperatures \emph{T}$_c$'s synthesized at various conditions of annealing temperature ($T_{an}$) and time. }
\begin{ruledtabular}
\renewcommand\arraystretch{1.5}
\begin{tabular}{ccccc}
 \textbf{No.} & \textbf{Ratio} &  \textbf{$T_{an}$ (K)} &  \textbf{time (days)} &  \textbf{\emph{T}$_c$ (K)} \\\hline
    7                  & 3                    & 605             & 10            & 4.3, 7.2  \\
  16                 & 3                    & 605             & 10            & 6.1, 7.2  \\
  27                 & 3                    & 605             & 10            & 7.2  \\
  43                 & 4                    & 623             & 3             & 7.2  \\
  52                 & 4                    & 615             & 3             & 3.5  \\
  50                 & 4     & {pestle}  & none  & 7.2  \\
  10                 & 3     & {pestle}  & none & 120  \\
\end{tabular}
\end{ruledtabular}
\end{table}

\begin{figure*}[htbp]
\includegraphics[width=2\columnwidth]{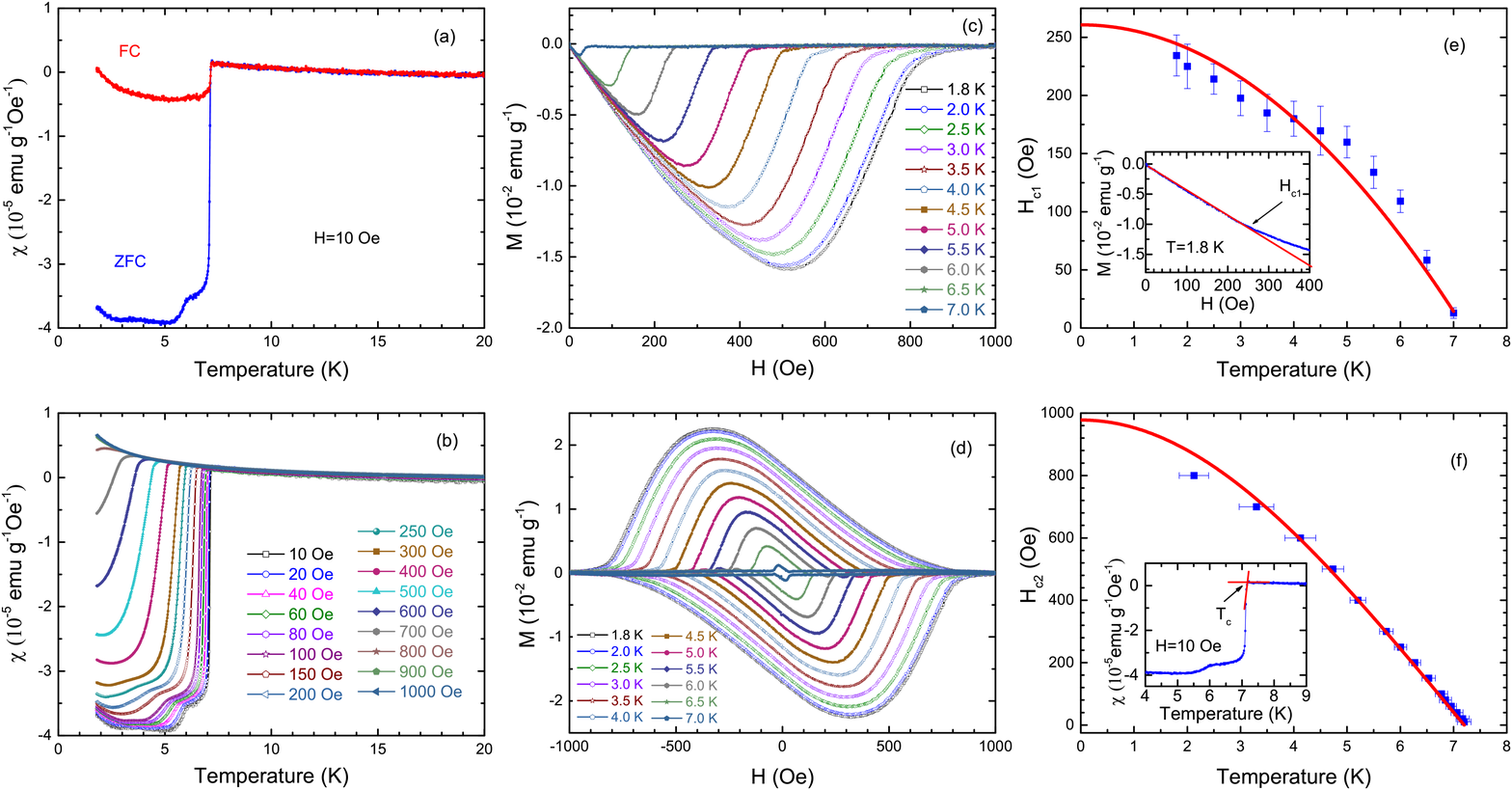}
\caption{(a) The temperature dependence of the magnetic susceptibility $\chi$ for potassium-doped \emph{p}-quaterphenyl (labeled by $\#$16) in the applied magnetic field of 10 Oe with field-cooling (FC) and zero-field cooling (ZFC). (b) The temperature dependence of the $\chi$ for potassium-doped \emph{p}-quaterphenyl measured at various magnetic fields up to 1000 Oe in the ZFC run. (c) The magnetic field dependence of the magnetization for potassium-doped \emph{p}-quaterphenyl at various temperatures in the superconducting state. (d) The magnetization hysteresis loop with scanning magnetic field along two opposite directions up to 1000 Oe measured at various temperatures in superconducting state. (e) The temperature dependence of the lower critical field \emph{H}$_{c1}$(T). The error bars represent estimated uncertainty in determining \emph{H}$_{c1}$. The inset shows the magnetic field dependence of the magnetization at the temperature of 1.8 K and the method for the determination of \emph{H}$_{c1}$. (f) The temperature dependence of the upper critical field \emph{H}$_{c2}$(T). The error bars represent the uncertainty in the rounding of the transition. The inset shows that \emph{T}$_c$ is defined on the curve of the temperature dependence of the magnetic susceptibility in the applied magnetic field of 10 Oe with the ZFC run.}
\end{figure*}

\begin{figure*}[tbp]
\includegraphics[width=2\columnwidth]{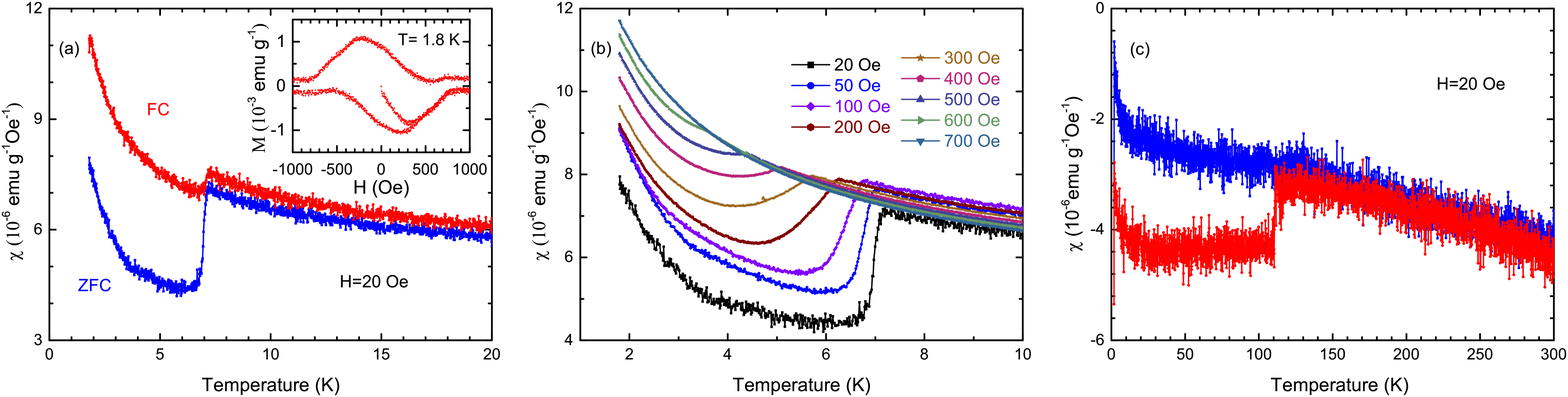}
\caption{(a) The temperature dependence of the magnetic susceptibility $\chi$ for potassium-doped \emph{p}-quaterphenyl (labeled by $\#$50) in the applied magnetic field of 10 Oe with the FC and ZFC runs. The inset shows the magnetization hysteresis loop with scanning magnetic field along two opposite directions up to 1000 Oe at the temperature of 1.8 K. (b) The temperature dependence of the magnetic susceptibility of potassium-doped \emph{p}-quaterphenyl (labeled by $\#$16) measured at various magnetic fields up to 700 Oe in the FC run. (c) The temperature dependence of the magnetic susceptibility $\chi$ for potassium-doped \emph{p}-quaterphenyl (labeled by $\#$10) in the applied magnetic field of 20 Oe with the FC and ZFC runs.}
\end{figure*}

The potassium-doped $p$-quaterphenyl samples were synthesized in diverse conditions. The paramagnetic behavior in the samples without superconductivity fit well with Curie-Weiss Law. Nevertheless, several samples, listed in Table I, present superconductivity. The primary superconducting phase with critical temperature of 7.2 K is repeated in the samples $\#$7, $\#$16, $\#$27, $\#$43 and $\#$50. The adding 4.3 K and 6.1 K superconducting phases accompany with the primary superconducting phase, respectively. Moreover, superconductivity of 7.2 K is obtained by pestling the mixtures of \emph{p}-quaterphenyl and potassium without annealing (labeled by $\#$50). Interestingly, another weak superconducting phase with a high \emph{T}$_c$ of 120 K is observed in the sample $\#$10.

Superconductivity of potassium-doped \emph{p}-quaterphenyl is characterized by magnetization measurements. Figure 1(a) displays the temperature dependence of the magnetic susceptibility $\chi$ in a low magnetic field of 10 Oe with the field-cooling (FC) and zero-field-cooling (ZFC) runs for the sample (labeled by $\#$16). The ZFC curve shows a sharp drop at the temperature of 7.2 K with a second drop at the temperature of 6.1 K, while a platform after the drop at the temperature of 7.2 K can be observed in the FC run. The magnetic measurements of this sample reveal two superconducting phases with critical temperatures of 6.1 K and 7.2 K, respectively. Supposing that the density $\delta$ of this sample is about 3 g/cm$^3$, we get the shielding fraction 4$\pi$$\chi$$\delta$=0.1488$\%$ from Fig. 1(a). The volume fraction of superconducting content is not very high due to the presence of impurities.

The obtained superconductivity of potassium-doped \emph{p}-quaterphenyl is further demonstrated by the curve of the temperature dependence of the $\chi$ measured at various magnetic fields. In Fig. 1(b), the $T$$_c$ gradually diminishes with increasing magnetic field since the superconducting fraction is suppressed by the applied magnetic field. The second 6.1 K superconducting phase is hardly observed at the magnetic field larger than 400 Oe, while the 7.1 K superconducting phase maintains until 1000 Oe.

The Meissner effect of this superconductor is further demonstrated by magnetization measurements. Figures 1(c) and 1(d) present the magnetization hysteresis loop up to 1000 Oe at various temperatures from 1.8 K to 7 K in the superconducting state, after removing the paramagnetic background (a straight line). In Fig. 1(c), the lower critical magnetic field \emph{H}$_{c1}$ is determined by the deviation from linearity in M versus H curves. The linearity region largens with decreasing temperature, meanwhile, the \emph{H}$_{c1}$ becomes higher along with lower temperature. Figure 1(d) represents the magnetization hysteresis loop with magnetic field along two opposite directions up to 1000 Oe at various temperature from 1.8 K to 7 K in superconducting state. The diamond-like shape of the magnetization hysteresis loop indicates that this sample is a typical type-$\amalg$ superconductor.

The $H$$_{c1}$ values at these selected temperatures are summarized in Fig. 1(e). The inset shows the method to determine $H$$_{c1}$ based on the deviation of the linear behavior at higher field. The zero-temperature extrapolated value of $H$$_{c1}$(0) is 261$\pm$9.6 Oe by means of calculating the empirical law $H$$_{c1}$($T$)/$H$$_{c1}$(0)=1-($T$/$T$$_c$)$^2$. Figure 1(f) shows the $H$$_{c2}$ versus $T$$_c$ curve, and the $T$$_c$ at applied filed is determined from the intercept of linear extrapolations from below and above the transition as shown in the inset. The calculated $H$$_{c2}$(0) at zero temperature is 978$\pm$7.5 Oe by using the Werthamer-Helfand-Hohenberg (WHH) formula\cite{WHH} $H$$_{c2}$(0)=0.693$\times$[-$d$$H$$_{c2}$($T$)/$d$$T$]$_{\emph{T}_c}$$\times$$T$$_c$, and 1092$\pm$39 Oe by fitting $H$$_{c2}$($T$) with the expression $H$$_{c2}$($T$)=$H$$_{c2}$(0)$\times$[1-($T$/$T$$_c$)$^2$]/[1+($T$/$T$$_c$)$^2$] based on the Gonzburg-Landau theory.

The sample $\#$50 was uncomplicatedly synthesised by just pestling the mixture of potassium and \emph{p}-quaterphenyl with a mole ratio of 4:1 without annealing. As shown in Fig. 2, the magnetization measurements provide the evidence of the existence of superconductivity in this sample. Figure 2(a) represents the temperature dependence of the $\chi$ in the applied magnetic field of 20 Oe with the FC and ZFC runs. The ZFC curve shows a sudden decrease at the temperature of 7.2 K, and a small upturn after the drop can be seen in the FC run. There is an obvious paramagnetic background accompanying with superconducting phase. The inset of Fig. 2(a) shows the magnetization hysteresis loop with magnetic field along two opposite directions up to 1000 Oe at 1.8 K after removing the paramagnetic background. The shape of this loop is a typical character for a superconductor. Figure 4(b) represents the temperature dependence of the $\chi$ measured at different magnetic fields up to 700 Oe in the ZFC processes. The superconducting fraction is suppressed by the applied magnetic fields. When the field is higher than 700 Oe, the superconducting transition is hardly observed. These features reveal that the sample $\#$50 is a new superconductor which is synthesised by just pestling without annealing.

In addition, a weak superconducting phase with a high \emph{T}$_c$ of 120 K is observed in the sample $\#$10. The sample $\#10$ was synthesised by pestling the mixture of potassium and \emph{p}-quaterphenyl with a mole ratio of 3:1 without annealing. Figure 2(c) displays the temperature dependence of the $\chi$ for the sample $\#$10 in the applied magnetic field of 20 Oe with the FC and ZFC processes. The $\chi$ versus \emph{T} curve exhibits a clear drop at the temperature of 120 K in the ZFC run, while the drop is too weak to be observed precisely in the FC run. This result supports the discovery of superconductivity in potassium-doped \emph{p}-terphenyl with $T$$_c$ of 123 K.\cite{terphenyl3} The 120 K superconducting phase may be the common feature in \emph{p}-oligophenyls with single C-C connected bond(s).

\begin{figure}[bp]
\includegraphics[width=1\columnwidth]{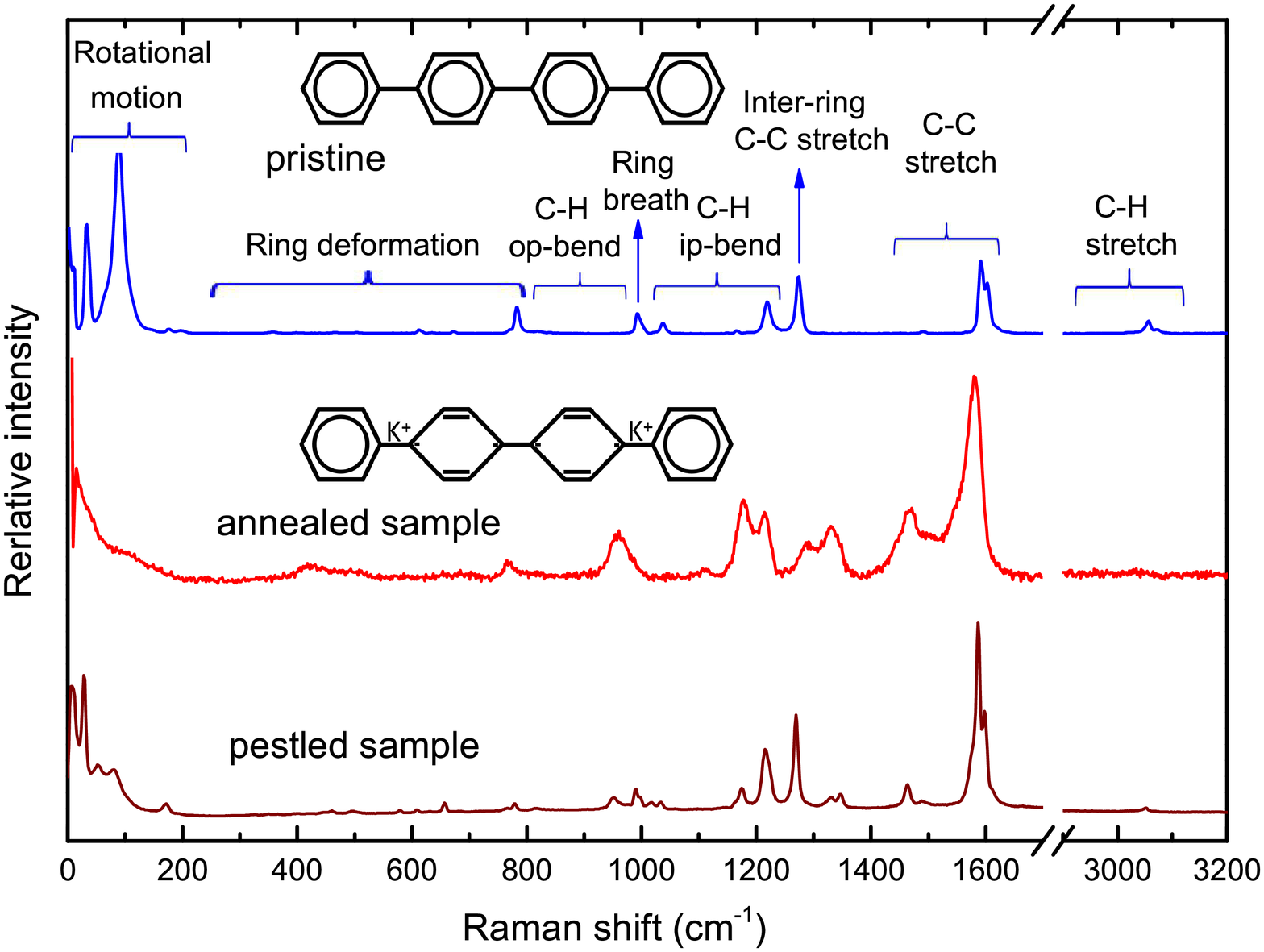}
\caption{Room-temperature Raman scattering spectra of pristine (blue line), annealed sample (red line), and pestled sample (brown line). The sticks in the upper horizontal axis give the peak positions of the vibrational modes in the pristine.}
\end{figure}

Figure 3 shows the Raman scattering spectra of the pristine, the annealed, and the pestled samples, respectively. The sticks over the pristine sample give the clarifications of the vibrational modes of the pristine $p$-quaterphenyl.\cite{mode} The Raman spectra of the annealed sample can be reduplicated in most annealed samples. This Raman spectrum in red line implies the formation of bipolarons, the scenario is consistent with previous works.\cite{bipolaron} According to the theory of bipolarons, two potassium atoms intercalate into the C-C bands between adjacent rings upon doping. Along the chain, the C-C bands between rings are reduced. In the inner-rings, the parallel bonds decrease whereas the inclined bonds increase. As a result, the inner-rings become quinoid. The molecule becomes nearly coplanar conformation and forms the conjugate conformation in the chain, and it leads to high intrachain mobilities of charge carriers such as bipolarons.\cite{coplanar} Contrast with the Raman spectra of pristine and annealed sample, the following five aspects reveal the formation of bipolarons. Firstly, the bipolaronic bands centered at 1581 cm$^{-1}$ appear after the merging of the strong bands at 1591 cm$^{-1}$ and 1601 cm$^{-1}$. They show downshifts in wavenumber due to the increase of the inclined C-C band lengths within the rings.\cite{bipolaron} Secondly, the new bipolaronic band at 1470 cm$^{-1}$ is associated with the vibrational mode of the C-H bend of external rings.\cite{bipolaron} Thirdly, the 1290 cm$^{-1}$ and 1332 cm$^{-1}$ bands in the annealed sample originate from the inter-ring C-C stretching mode at 1274 cm$^{-1}$ band of the pristine. The upshifts in wavenumber reflect the decrease of length in the C-C bonds between rings. Fourthly, the two bipolaronic bands at 1179 cm$^{-1}$ and 1215 cm$^{-1}$ correspond to the 1218 cm$^{-1}$ band in the pristine. Finally, the new bipolarnic band at 961 cm$^{-1}$ arises from the in-plane ring bend. Furthermore, in the pestled sample, it can be obviously seen the coexistence of the pristine and bipolarons. Nevertheless, the intensity of the phonon modes in pristine is stronger than those of bipolarons. The strong bands at around 1600 cm$^{-1}$ are the overlap with the pristine and the bipolaron. The new band at 1465 cm$^{-1}$ in the pestled sample is homologous to the band at 1470 cm$^{-1}$ in annealed sample. The band at 1274 cm$^{-1}$ in pristine evolves into bipolaronic bands at 1330 cm$^{-1}$ and 1347 cm$^{-1}$ in pestled sample, and the 1274 cm$^{-1}$ band still remains in pestled sample as the component of pristine. The 1174 cm$^{-1}$ band in pestled sample arises from the 1218 cm$^{-1}$ band in pristine. Moreover, in the pestled sample parts of the bipolaronic bands are overlapped by the signal of pristine. Admittedly, the Raman spectra provide the evidences of the formation of bipolarons in both pestled and annealed samples.

$p$-Oligophenyls without the zigzag-like and armchair structures are different from polycyclic aromatic hydrocarbons such as picene,\cite{picene,res} phenanthrene,\cite{threne,huang}  coronene,\cite{43} and 1,2:8,9-dibenzopentacene.\cite{pentacene} The current work together with our recent experimental findings\cite{terphenyl1,terphenyl2,terphenyl3,quinquephenyl,bipyridine} adds a new family of organic superconductors connected by single C-C linked bond(s).

\section{CONCLUSIONS}

Encouraged by the discovery of superconductivity in potassium-doped \emph{p}-terphenyl, we examine such a possibility in \emph{p}-quaterphenyl with one more phenyl ring than \emph{p}-terphenyl based on the consideration of that the conductivity increases with increasing chain length in poly-\emph{p}-phenylene family. The samples synthesised in various conditions of annealing or just pestling. The Meissner effect with critical temperatures ranging from 3.5 K to 120 K is observed by the magnetic susceptibility measurements. The superconducting phase of 7.2 K can be reduplicated in several samples with superconductivity and seems to be the most stable phase. Comparing with previous reports about superconductivity in potassium-doped \emph{p}-terphenyl, $p$-quinquephenyl, and 2,2$'$-bipyridine, the primary 7.2 K superconducting phase maybe the common character in chain linking molecules. Analogously, the weak 120 K superconducting phase in potassium-doped \emph{p}-quaterphenyl is corresponding to the superconducting phase with \emph{T}$_c$ above 120 K in potassium-doped \emph{p}-terphenyl. In both the annealed and pestled samples with superconductivity, the bipolaronic character is observed by the Raman scattering measurements. The discussion of that the mechanism of superconductivity whether from bipolarons remains a subject for further researches. The realization of superconductivity in potassium-doped \emph{p}-quaterphenyl supports the discovery of superconductivity in potassium-doped \emph{p}-terphenyl and adds a new leaguer of superconductors in $p$-oligophenyl family. The innovation of synthetic method by just pestling without annealing opens up a new royal road to synthesise organic superconductors.

We thank Guo-Hua Zhong, Hai-Qing Lin, Yun Gao, and Zhong-Bing Huang for valuable discussions.


\begin{references}

\bibitem{predict1} V. L. Ginzburg, High-temperature superconductivity-dream or reality? Sov. Phys. Usp. \textbf{19}, 174 (1976).

\bibitem{predict2} W. A. Little, Possibility of synthesizing an organic superconductor. Phys. Rev. \textbf{134}, A1416 (1964).

\bibitem{polymer} A. J. Heeger, S. Kivelson, J. R. Schrieffer, and W. P. Su, Solitons in conducting polymers. Rev. Mod. Phys. \textbf{60}, 781 (1988).

\bibitem{PPP1} L. W. Shacklette, R. R. Chance, D. M. Ivory, G. G. Miller, and R. H. Baughfnan, Electrical and optical properties of highly conducting charge-transfer complexes of poly ($p$-phenylene) Synth. Met. \textbf{1}, 307 (1979).

\bibitem{PPP2} P. W. Anderson, Model for the electronic structure of amorphous semiconductors. Phys. Rev. Lett. \textbf{34}, 953 (1975).

\bibitem{PPP3} P. Kuivalainen, H. Stubb, H. Isotalo, P. YliLahti, and C. Holmstr\"{�� o}m, Electrical and optical properties of FeCl$_3$-doped polyparaphenylene [($p$-C$_6$H$_4$)$_x$]. Phys. Rev. B \textbf{31}, 7900 (1985).

\bibitem{conduct} D. M. Ivory, G. G. Miller, J. M. Sowa, L. W. Shacklette, R. R. Chance, and R. H. Baughman, Highly conducting charge-transfer complexes of poly($p$-phenylene). J. Chem. Phys. \textbf{71}, 1506 (1979).

\bibitem{upconduct} E. E. Havinga, and L. W. Van Horssen, Dependence of the electrical conductivity of heavily-doped poly-$p$-phenylenes on the chain length. Synthetic Metals \textbf{16}, 55 (1986).

\bibitem{quinoid} J. L. Br��das, Relationship between band gap and bond length alternation in organic conjugated polymers. J. Chem. Phys. \textbf{82}, 3808 (1985).

\bibitem{terphenyl1} R. S. Wang, Y. Gao, Z. B. Huang, and X. J. Chen, Superconductivity in $p$-terphenyl. arXiv:1703.05803.

\bibitem{terphenyl2} R. S. Wang, Y. Gao, Z. B. Huang, and X. J. Chen, Superconductivity at 43 K in a single C-C bond linked terphenyl. arXiv:1703.05804.

\bibitem{terphenyl3} R. S. Wang, Y. Gao, Z. B. Huang, and X. J. Chen, Superconductivity above 120 kelvin in a chain link molecule. arXiv:1703.06641.

\bibitem{benzene} G. H. Zhong, X. J. Chen, and H. Q. Lin, Prediction of superconductivity in potassium-doped benzene. arXiv:1501.00240.

\bibitem{structure} G. H. Zhong, X. H. Wang, R. S. Wang, J. X. Han, C. Zhang, X. J. Chen, and H. Q. Lin, Structural and bonding character of potassium-doped $p$-terphenyl superconductors. arXiv:1706.03965.

\bibitem{quinquephenyl} G. Huang, R. S. Wang, and X. J. Chen, Observation of Meissner effect in potassium-doped $p$-quinquephenyl. arXiv:1801.06324.

\bibitem{bipyridine} K. Zhang, R. S. Wang, A. J. Qin, and X. J. Chen, superconductivity in potassium-doped 2,2$'$-bipyridine. arXiv:1801.06320.

\bibitem{ARPES} H. X. Li, X. Q. Zhou, S. Parham, T. Nummy, J. Griffith, K. Gordon, E. L. Chronister, and D. S. Dessau, Spectroscopic evidence of pairing gaps to 60 Kelvin or above in surface-doped $p$-terphenyl crystals. arXiv:1704.04230.

\bibitem{FengDL} M. Q. Ren, W. Chen, Q. Liu, C. Chen, Y. J. Qiao, Y. J. Chen, G. Zhou, T. Zhang, Y. J. Yan, and D. L. Feng, Observation of novel gapped phases in potassium doped single layer $p$-terphenyl on Au (111). arXiv:1705.09901.

\bibitem{india} P. Neha, V. Sahu, and S. Patnaik. Facile synthesis of potassium intercalated $p$-terphenyl and signatures of a possible high $T$$_c$ phase. arXiv:1712.01766.

\bibitem{DOS} R. M. Geilhufe, S. S. Borysov, D. Kalpakchi, and A. V. Balatsky, Towards novel organic high-$T$$_c$ superconductors: data mining using density of states similarity search. arXiv:1709.03151. 

\bibitem{twoband} M. Fabrizio, T. Qin, S. S. Naghavi, and E. Tosatti, Two-band s$_{\pm}$ strongly correlated superconductivity in K$_3$ p-terphenyl? arXiv:1705.05066.

\bibitem{fano} M. V. Mazziotti, A. Valletta, G. Campi, D. Innocenti, A. Perali, and A. Bianconi, Possible Fano resonance for high $T$$_c$ multi-gap superconductivity in p-terphenyl doped by K at the Lifshitz transition. arXiv:1705.09690.

\bibitem{WEN} W. H. Liu, H. Lin, R. Z. Kang, Y. Zhang, X. Y. Zhu, and H. H. Wen, Magnetization of potassium doped $p$-terphenyl and $p$-quaterphenyl by high pressure synthesis. Phys. Rev. B \textbf{96}, 224501 (2017).

\bibitem{OTFT} D. J. Gundlach, Y. Y. Lin, T. N. Jackson, and D. G. Schlom, Oligophenyl-based organic thin film transistors. Appl. Phys. Lett. \textbf{71}, 3853 (1997).

\bibitem{OLED} C. Hosokawa, H. Higashi, and T. Kusumoto, Novel structure of organic electroluminescence cells with conjugated oligomers. Appl. Phys. Lett. \textbf{62}, 3238 (1993).

\bibitem{laser} F. Quochi, Random lasers based on organic epitaxial nanofibers. J. Opt. \textbf{12.2}, 024003 (2010).

\bibitem{optical} G. Leising, S. Tasch, F. Meghdadi, L. Athouel, G. Froyer, and U. Scherf, Blue electroluminescence with ladder-type poly($para$-phenylene) and $para$-hexaphenyl. Synth. Met. \textbf{81}, 185 (1996).

\bibitem{crystal} Y. Sakamoto, T. Suzuki, A. Miura, H. Fujikawa, S. Tokito, and Y. Taga, Synthesis, characterization, and electron-transport property of perfluorinated phenylene dendrimers. J. Amer. Chem. Soc. \textbf{122.8}, 1832 (2000).

\bibitem{carrier} S. Kania, W. Mycielski, and A. Lipi��ski,  Charge carrier transport in oriented $p$-quaterphenyl layers. Thin Solid Films \textbf{61.2}, 229 (1979) .

\bibitem{derivative} J. P. Lu, K. J. Miyatake, A. R. Hlil, and A. S. Hay, Novel soluble and fluorescent poly(arylene ether)s containing $p$-quaterphenyl, 2,5-Bis(4-phenylphenyl)oxadiazole, or 2,5-bis(4-phenylphenyl)triazole groups. Macromolecules \textbf{34}, 17 (2001).

\bibitem{anneal} A. A. A. Darwish, Effect of annealing on structural, electrical and optical properties of $p$-quaterphenyl thin films. Infrared Phys. Technol. \textbf{82}, 96 (2017).

\bibitem{temperature} L. Athouel, R. Resel, N. Koch, G. Froyer, and G. Lelsing, Orientation of molecules in phenylene oligomer thin films: influence of the substrate temperature. Synthetic Metals, \textbf{101}, 627 (1999).

\bibitem{magnet} T. Mori, K. Mori, T. Mizutani, Effect of magnetic field on growth of $p$-quaterphenyl thin films. Thin Solid Films \textbf{366}, 279 (2000).

\bibitem{pattern} G. Hlawacek, C. Teichert, S. M\"{u}llegger, R. Resel, and A. Winkler, Pattern formation in $para$-quaterphenyl film growth on gold substrates. Synthetic Metals, \textbf{146.3}, 383 (2004).

\bibitem{WHH} N. R. Werthamer, E. Helfand, and P. C.Hohenberg, Temperature and purity dependence of the superconducting critical field, \emph{H}$_{c2}$ III. electron spin and spin-orbit effects. Phys. Rev. \textbf{147}, 295 (1966).

\bibitem{mode} K. Honda and Y. Furukawa, Conformational analysis of \emph{p}-terphenyl by vibrational spectroscopy and density functional theory calculations. J. Mol. Struct. \textbf{11}, 735 (2005).

\bibitem{bipolaron} M. Dubois, G. Froyer, G. Louarn, and D. Billaud, Raman spectroelectrochemical study of sodium intercalation into poly(\emph{p}-phenylene). Spectrochim. Acta, Part A \textbf{59}, 1849 (2003).

\bibitem{coplanar} J. L. Br\'{e}das, G. B. Street, B. Th\'{e}mans, and J. M. Andr\'{e}, Organic polymers based on aromatic rings (polyparaphenylene, polypyrrole, polythiophene): evolution of the electronic properties as a function of the torsion angle between adjacent rings. J. Chem. Phys. \textbf{83}, 1323 (1985).

\bibitem{picene} R. Mitsuhashi, Y. Suzuki, Y. Yamanari, H. Mitamura, T. Kambe, N. Ikeda, H. Okamoto, A. Fujiwara, M. Yamaji, N. Kawasaki, Y. Maniwa, and Y. Kubozono, Superconductivity in alkali-metal-doped picene. Nature \textbf{464}, 76 (2010).

\bibitem{res} K. Teranishi, X. He, Y. Sakai, M. Izumi, H. Goto, R. Eguchi, and Y. Kubozono, Observation of zero resistivity in K-doped picene, Phys. Rev. B \textbf{87}, 060505 (2013).

\bibitem{threne} X. F. Wang, R. H. Liu, Z. Gui, Y. L. Xie, Y. J. Yan, J. J. Ying, X. G. Luo, and X. H. Chen, Superconductivity at 5 K in alkali-metal-doped phenanthrene. Nat. Commun. \textbf{2}, 507 (2011).

\bibitem{huang} Q. W. Huang, G. H. Zhong, J. Zhang, X. M. Zhao, C. Zhang, H. Q. Lin, and X. J. Chen, Constraint on the potassium content for the superconductivity of potassium-intercalated phenanthrene. J. Chem. Phys. \textbf{140}, 114301 (2014).

\bibitem{43} Y. Kubozono, H. Mitamura, X. Lee, X. He, Y. Yamanari, Y. Takahashi, and T. Kambe, Metal-intercalated aromatic hydrocarbons: a new class of carbon-based superconductors, Phys. Chem. Chem. Phys. \textbf{13}, 16476 (2011).

\bibitem{pentacene}  M. Xue, T. Cao, D. Wang, Y. Wu, H. Yang, X. Dong, J. He, F. Li, and G. F. Chen, Superconductivity above 30 K in alkali-metal-doped hydrocarbon. Sci. Rep. \textbf{2}, 389 (2012).

\end{references}
\end{document}